\documentclass[mathleft
]{an}
\usepackage{graphicx}
\usepackage{times}

\begin{document}

% The following seven commands are intended for editorial usage and should be ignored by
% the author(s).
\Pagespan{789}{}% Document's page range. 
% If second parameter is left empty, the last page is computed automatically.
\Yearpublication{2006}%
\Yearsubmission{2005}%
\Month{11}%   
\Volume{999}%  
\Issue{88}% 
% \DOI{This.is/not.aDOI}% 

\title{Broadband short term X-ray variability of the quasar PDS 456}

\author{G. A. Matzeu\inst{1}\fnmsep\thanks{Corresponding author:
  \email{g.matzeu@keele.ac.uk}\newline}
%Example 
%for footnote, note the usage of the \texttt{fnmsep}
%command as separator between institute number and footnote mark} 
,  J. N. Reeves\inst{1,2}, E. Nardini\inst{1}, V. Braito\inst{3}, M. T. Costa\inst{1}, F. Tombesi\inst{4,5}, J Gofford\inst{1}
}
\titlerunning{Short term variability in PDS 456}
\authorrunning{G. A. Matzeu et al.}
\institute{
Astrophysics Group, School of Physical and Geographical Sciences, Keele University, Keele, Staffordshire ST5 5BG, UK
\and 
Center for Space Science and Technology, University of Maryland Baltimore County, 1000 Hilltop Circle, Baltimore, MD 21250, USA
\and 
INAF – Osservatorio Astronomico di Brera, Via Bianchi 46, I-23807 Merate (LC), Italy
\and
X-ray Astrophysics Laboratory, NASA/Goddard Space Flight Center, Greenbelt, MD, 20771, USA
\and
Department of Astronomy and CRESST, University of Maryland, College Park, MD, 20742, USA}

\received{30 May 2005}
\accepted{11 Nov 2005}
\publonline{later}


\abstract{We present a detailed analysis of a recent $500$ ks net exposure \textit{Suzaku} observation, carried out in 2013, of the nearby ($z=0.184$) luminous (L$_{\rm bol}\sim10^{47}$ erg s$^{-1}$) quasar PDS 456 in which the X-ray flux was unusually low. The short term X-ray spectral variability has been interpreted in terms of variable absorption and/or intrinsic continuum changes. In the former scenario, the spectral variability is due to variable covering factors of two regions of partially covering absorbers. We find that these absorbers are characterised by an outflow velocity comparable to that of the highly ionised wind, i.e. $\sim0.25$ c, at the $99.9\%$ $(3.26\sigma)$ confidence level. This suggests that the partially absorbing clouds may be the denser clumpy part of the inhomogeneous wind. Following an obscuration event we obtained a direct estimate of the size of the X-ray emitting region, to be not larger than $20~R_{\rm g}$ in PDS 456.}

\maketitle

\section{Introduction}
It has now being widely recognised that outflows are an indispensable component in the overall understanding of Active Galactic Nuclei (AGN). These winds are believed to occur as a result of the accretion process (King 2003) hence providing a link between the black hole mass and the velocity dispersion of the stars in the bulge of a galaxy, such as seen with the $M-\sigma$ relation (Ferrarese $\&$ Merritt 2000; Gebhardt et al. 2000). Since the last decade, a number of massive and high velocity outflows have been observed in luminous AGN (Reeves et al. 2009; Gofford et al. 2013) through the presence of blueshifted resonance Fe K absorption lines at rest-frame energies above $>7$ keV. The importance of these winds is supported by their frequent detection, as they are observed in the X-ray spectra of approximately $40\%$ of AGN (Tombesi et al 2010), suggesting that their geometry is characterised by a wide solid angle. This  was recently confirmed in the quasar PDS 456 by Nardini et al. (2015, N15 hereafter). The mechanical power produced by these high velocity outflows is possibly exceeding the $0.5-5\%$ of the bolometric luminosity $L_{bol}$ required for a significant feedback contribution in the co-evolution of the AGN host galaxy (Hopkins \& Elvis 2010). AGN are also know for their spectral variability; in particular when it is present in the X-rays, it is possible to probe the inner regions near the super-massive black holes. The AGN X-ray variability could be characterised by both intrinsic fluctuating spectral and/or temporal behaviour or in part by the presence of absorbing gas in the Line-Of-Sight (LOS). The latter case may be seen as changes in the covering fraction of a partial covering absorber (Turner et al. 2011).
\\  
The luminous radio-quiet quasar PDS 456 is located at a red-shift of $z=0.184$ (Torres et al. 1997) and it has a de-reddened absolute magnitude of M$_B\sim-27$ and a bolometric luminosity of $L_{bol}=10^{47}$ erg s$^{-1}$ (Reeves et al. 2000). It is comparable in luminosity to the radio-loud quasar 3C 273, making it the most luminous quasar in the local Universe. Such a high luminosity is more typical of quasars at red-shift $z=2-3$, considered the peak of the quasar epoch. In 2001, a short ($\sim40$ ks) observation carried out with \textit{XMM-Newton} detected a strong absorption trough, above $7$ keV, possibly attributed to the highly ionised iron K-shell feature with an associated velocity outflow of the order of $v_{\rm w}\ga 0.1$ c (Reeves et al. 2003). A later $190$ ks \textit{Suzaku} observation carried out in 2007, confirmed the evidence of a fast outflow, revealing two highly significant absorption lines at $9.08$ and $9.66$ keV in the quasar rest frame where no strong atomic transitions are expected. The association of these lines to the nearest expected strong line, the iron XXVI Ly $-\alpha$ doublet transition at $6.97$ keV, implied an outflow velocity of $\sim0.30$ c (Reeves et al 2009). Few years later, during a series of five simultaneous observations with \textit{XMM-Newton} and \textit{NuSTAR} in $2013-2014$, N15 resolved a fast ($\sim0.25$ c) P-Cygni like profile at iron K, showing that the absorption arise from a wide angle accretion disc wind. PDS 456 has now a proven track record of strong spectral variability over the last decade of observations, likely due to absorption and/or intrinsic continuum variations. 
\\
In this work we present a long ($\sim1$ Ms duration) 2013 \textit{Suzaku} campaign carried out in order to determine the time-scales through which both the X-ray absorption and continuum variations occur by directly measuring the absorber behaviour on time-scales of approximately tens of ks (corresponding to a light-crossing time of a few $R_{\rm g}$ for $M_{BH}\sim10^{9}M_{\odot}$). In this work we investigate the broad-band continuum and absorption variability over the course of the 2013 observations.

\section{Broad-Band Spectral Analysis}

\begin{figure}[h!]
\includegraphics[width=70mm,height=70mm]{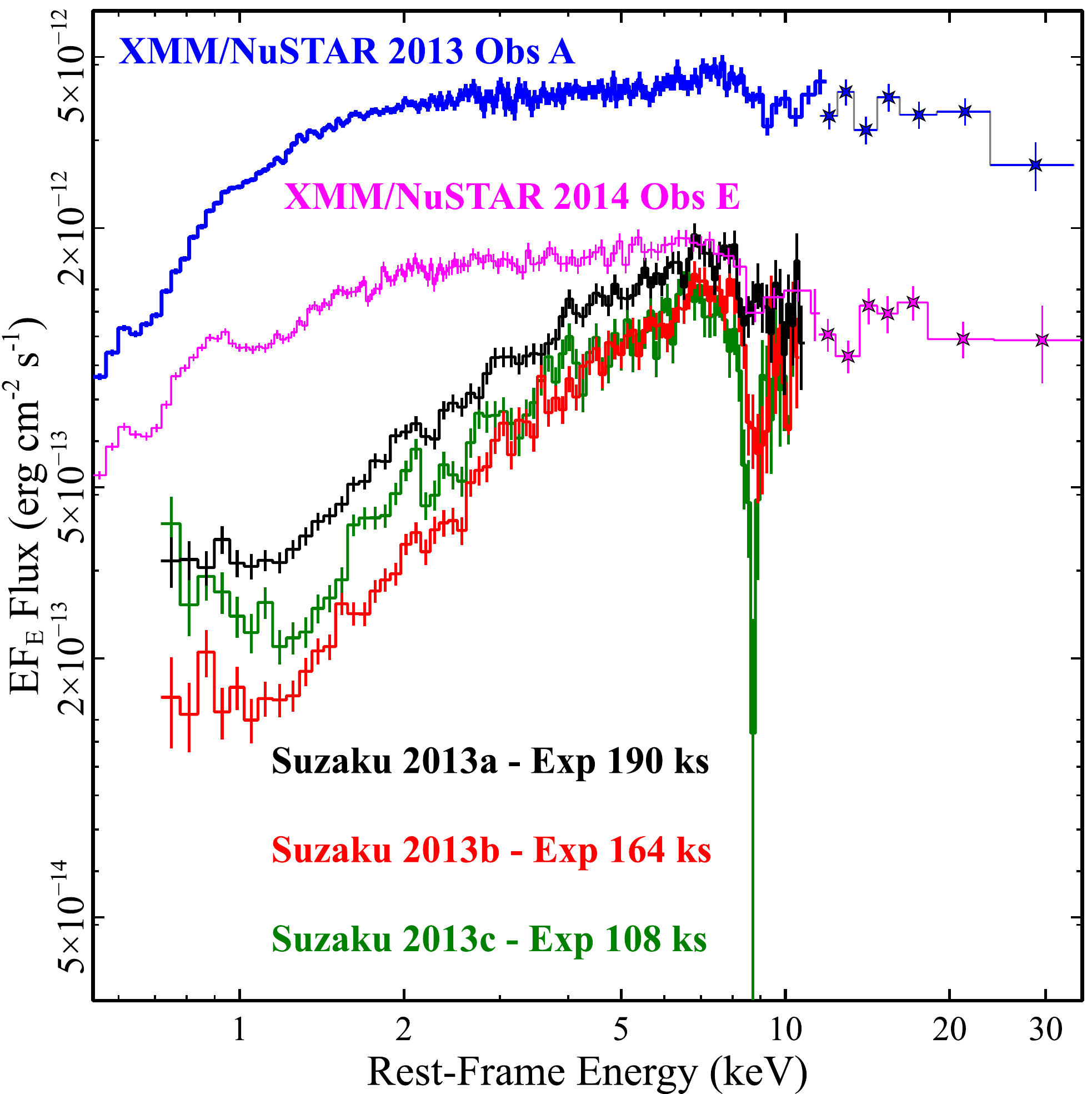}
\caption{The $0.5-40$ keV mean spectra of {\textit{XMM-Newton}}/{\textit{NuSTAR}} 2013 obs A (blue), 2014 obs E (magenta). The $0.7-10$ keV, Sequence 1 - 2013a (black), Sequence 2 - 2013b (red) and Sequence 3 - 2013c (green) of PDS 456 observed with \textit{Suzaku} showing the long term spectral changes in the soft band and in the Fe K region. The \textit{Suzaku} observation in 2013 caught PDS 456 in an unusually low flux state compared to the highest (Obs A) and lowest (ObsE) flux spectra of the later {\textit{XMM-Newton}}/{\textit{NuSTAR}} campaign.}
\label{pds_xmm_nu_suz_proceeding}
\end{figure}

The 2013 \textit{Suzaku} campaign caught PDS 456 in an unusually low flux, compared to the earlier 2007 and 2011 observations and also to the later XMM-Newton and \textit{NuSTAR} campaign, carried out in August 2013/February 2014. Fig.~\ref{pds_xmm_nu_suz_proceeding} shows the comparison between the flux spectra (unfolded through the instrumental response, versus a simple $\Gamma=2$ power-law, and not corrected for Galactic absorption) from the 2013 \textit{Suzaku} sequences to the lowest (Obs E) and the highest (Obs A) flux of the five \textit{XMM-Netwon/NuSTAR} sequences (N15). It follows that this 2013 campaign provides a unique opportunity to study PDS 456 in an extended low flux state in great detail. Before probing the short-term variability, we first parameterised the 2013 broad-band continuum spectra from the three \textit{Suzaku} sequences. We found that two layers of partial covering absorbers were required in order to account for the spectral curvature present in all three 2013 \textit{Suzaku} spectra (Matzeu et al. 2016).

\subsection{Time-Sliced Lightcurves}

\begin{figure}[h!]
\includegraphics[width=80mm,height=40mm]{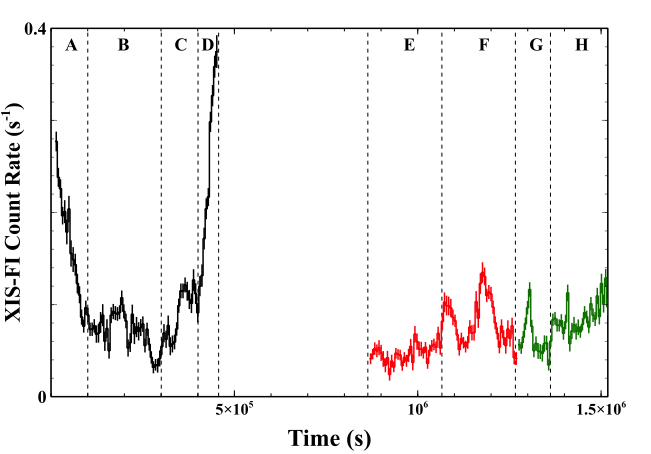}
\caption{XIS Front Illuminated (Broad-band light-curve ($0.5-10$ keV) including sequences 2013a (black), sequence 2013b (red) and sequence 2013c (green). The dashed vertical lines illustrate the size of the time-scale of the eight slices. Note the strong flare between $400-450$ ks in segment D.}
\label{fig:pds456_suz_2013_lc}
\end{figure}

Fig.~\ref{fig:pds456_suz_2013_lc} shows the overall light-curves of the 2013 \textit{Suzaku} campaign strongly indicating variability of the X-ray flux in PDS 456 on short time scales. A prominent flare is detected, with the flux increasing by a factor of $ \sim 4\times$ between $400-450$ ks during the observation in sequence 2013a followed by several smaller flares towards the second half of sequence 2013b. Guided by the visual properties of the overall light-curve the spectra were divided into eight time resolved slices (see Fig.~\ref{fig:pds456_suz_2013_lc}), taking into consideration the width of each slice and the number of counts in it. Note for PDS 456 with $M_{\rm BH}\sim10^{9}M_{\odot}$, a variability time-scale of $\sim100$ ks corresponds to a light-crossing distance of $\sim20$ R$_{\rm g}$ (where 1 R$_{\rm g} = {GM_{\rm BH}/c^2}$). It follows that within the first four slices (A - D), corresponding to 2013a observation, the decline of an initial flare is traced (slice A) followed by a quiescent period (slice B) together with the initial onset of the large flare (slice C) and the subsequent flare itself (slice D) of $\sim50$ ks in duration. In the remaining four slices (E - H), corresponding to both 2013b and 2013c observations, the Fe K absorption feature becomes significant (slice E), progressing in strength (slice F), reaching maximum depth in slice G and then a hint of recovering in slice H. All the spectral analysis and model fitting in this work are performed with XSPEC v 12.8.2 (Arnaud 1996)

\subsection{Partial Covering Variability}

Here we investigated models where X-ray photons are reprocessed through compact absorbing clouds, that partially absorb the X-ray emission allowing just a fraction ($1-f_{\rm{cov}}$) to emerge unattenuated. The typical size-scale of these absorbing clouds is comparable to the X-ray emitting region (i.e. tens of $R_{g}$). In this model, for simplicity, the partial covering column densities are kept constant within the slices; instead the spectral variability across the observation are due to variations in the covering fractions $ f_{\rm cov} $. For the partial coverer, the column density of the higher column region is found to be 
$\log(N_{\rm H,\rm high}/{\rm cm^{-2}})=23.3\pm0.1$, whilst the lower column zone has $\log(N_{\rm H,\rm low}/{\rm cm^{-2}})=22.3\pm0.1$. For the high column zone the covering fraction reaches its minimum during the flare, slice D ($f_{cov,\rm high}<0.38$) and increases to its maximum value ($f_{cov,\rm high}=0.60_{-0.06}^{+0.05}$) in slice E. In the lower column absorber, the minimum value is at $f_{cov,\rm low}=0.59_{-0.02}^{+0.07}$ in slice C increasing towards its maximum value $f_{cov,\rm low}=0.81\pm0.03$ in slice D. By examining these changes related to flux (Matzeu et al. 2016), we observe a clear correlation between the high column covering fraction variability and the observed flux in both $0.5-1$ keV and $2-10$ keV bands; on the other hand it appears that the low column covering fraction variability and flux anti-correlate, in most part of the observation in both energy bands. This trend may suggest that the high column covering fractions possibly account for most of the spectral variability, compared to the low column.

\subsubsection{Properties of the Partial Covering And Location Within the Outflow}

\begin{figure}[h!]
\begin{center}
\includegraphics[scale=0.45]{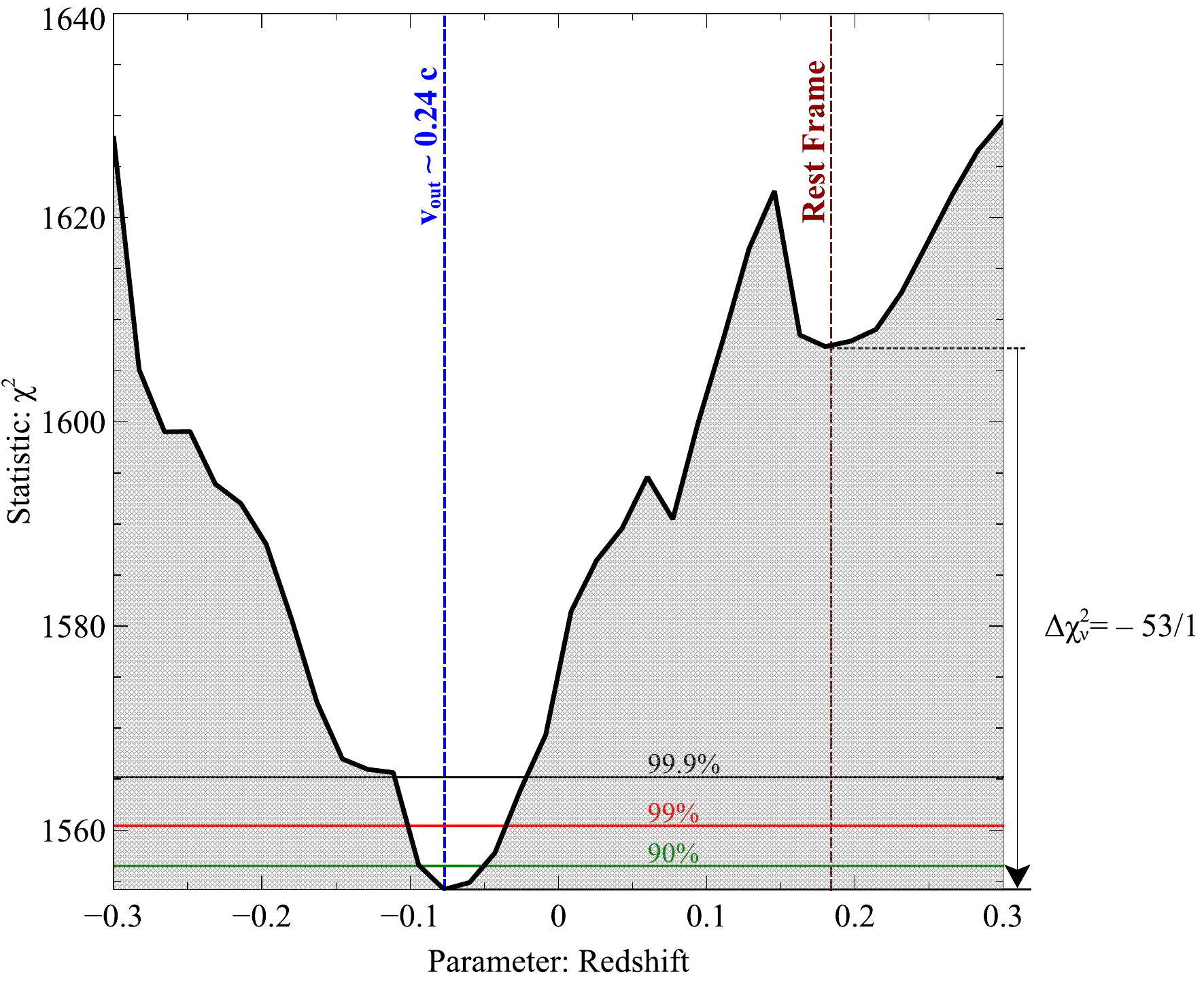}	
%%
%\vspace{100mm}
%%
\caption{One dimensional contour plot of the $\chi^2$ against the partially ionised partial covering red-shift parameter in all the slices. A local minimum at about the quasar rest frame (i.e. $z=0.184$) is also clearly visible. However by adopting a more complex model the global minima strongly suggests that the partial covering layers are outflowing at velocity comparable to that of the fully covering ionised wind i.e.$v_{\rm w}\sim0.24$ c constrained with a statistical improvement, in respect to the rest-frame velocity, of $\Delta\chi_{\nu}^{2}=-53/1$.}
\label{fig:pds456_2013_cont_ionpc_z}
\end{center}	
\end{figure}

From the previous section we find that we cannot explain the overall short-term spectral variability without evoking a variable partial coverer. Thus a reasonable question that follows up this result is:- what is the partial coverer associated to ?
\\
It may be plausible that it is indeed associated with the outflow and to be more specific it may be the less ionised and more dense part of the wind. Thus we tested a model scenario where we kept both the high and low column partial coverer red-shift parameters (as a proxy for the its velocity i.e.  $v_{\rm pc}$) constant between all the eight slices. The resulting fit suggested that the gas is indeed outflowing with a velocity consistent with the velocity of the highly ionised gas. In Fig.~\ref{fig:pds456_2013_cont_ionpc_z} a confidence contour plot of the $\chi^2$ against the partial covering red-shift parameter tied between the all the slices is shown. A local minimum at about the quasar rest frame (i.e. $z = 0.184$) is clearly visible, however the fit statistic is worse by $\Delta\chi^{2}/\Delta\nu=53/1$ compared to the best-fit case. The global minimum in Fig.~\ref{fig:pds456_2013_cont_ionpc_z} indicates that $v_{\rm pc}\sim0.24$ c outflow velocity is preferred over a systemic velocity at the $\sigma=7.3$ ($>99.99\%$) confidence level. This may indicate that the two partially covering zones are the least ionised component of the \textit{same} fast $(v_{\rm pc}\sim v_{\rm w}\sim0.24~c)$ wind.

\section{Discussion}

In this work we found that the observed short-term spectral variability can be explained by the combination of both variable partial covering absorbers and a variable intrinsic continuum also investigated in Matzeu et al. (2016); although the first scenario may be the dominant cause in this observation. Furthermore by investigating in more detail the properties of the partial covering absorbers, we found that its outflow velocity is comparable to that of the highly ionised wind (i.e. $v_{\rm pc}\sim v_{\rm w}\sim0.24~c$) in both cases where the absorber is either neutral or mildly ionised. This implies that the partial covering may be the less ionised component of the highly ionised outflow.

\subsection{Short-Term X-ray Spectral Variability Of The Fe K Absorption Feature}

Across the observation, the Equivalent Width (EW) and the column density $N_{\rm H}$ of the Fe K absorption feature increased by a factor of $\sim10$. These variations are possibly caused by a transit of a dense cloud or stream of highly ionised material moving across the line of sight as part of an inhomogeneous wind. The linear size of the transiting clump, across the LOS, can be estimated from $\Delta R=v_{\rm w}\Delta t$ by assuming that any transverse velocity is comparable to the wind velocity, i.e., $v_{\rm K}\sim v_{\rm w}$. The choice of $\Delta t=400$ ks corresponds to the time between the initial (statistically significant) onset of the absorption profile in slice E ($\sim3\sigma$) and when it reaches maximum depth in slice G ($>5\sigma$). By adopting $v_{\rm w}\sim0.25$ c, we can estimate that $\Delta R \sim 3\times10^{15}$ cm $\sim20 R_g$ in PDS 456. 

\subsubsection{Constraint On The X-ray Emitting Region}

\begin{figure}[h!]
\includegraphics[width=70mm,height=45mm]{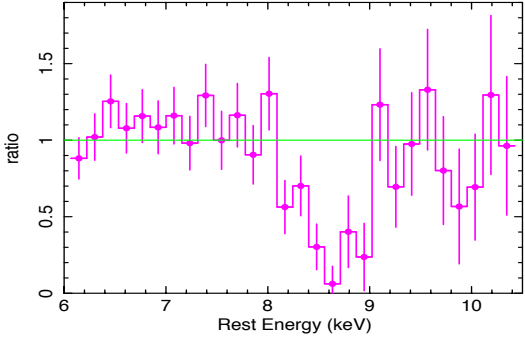}
\caption{Zoom-in of the ratio plot of slice G approximately binned at Full Width Half Maximum (FWHM) resolution. The continuum flux reaches practically zero flux at $\sim8.6$ keV, indicating the highly ionised absorber may be fully covering the X-ray emitter.} 
\label{fig:ratio_G_edit}
\end{figure}

For extra clarity we plotted a zoom-in of slice G in Fig.~\ref{fig:ratio_G_edit} showing the Fe K absorption profile where the flux at the absorption line centroid is consistent with zero. This suggests that the highly ionised clump is likely fully covering the X-ray source. Therefore we can obtain an independent constraint on the size of the X-ray emitting region in PDS 456 which cannot be larger than $\sim20 R_g$ for the absorber to fully cover the X-ray emission region.

\subsection{Can The Flare Drive The Outflow ?}
 
One question arising is whether the strong flare emission, shown in Fig.~\ref{fig:pds456_suz_2013_lc}, could radiatively power the outflow observed in the latter part of the observation. From the photon momentum transfer to the wind it follows:-

\begin{equation}
\dot{p}_{\rm w}=\dot{M}_{\rm w} v_{\rm w}=\tau\frac{L_{\rm flare}}{c},
\end{equation}
where $\dot{p}_{\rm w},\dot{M}_{\rm w}$ and $v_{\rm w}$ are the momentum rate, mass outflow rate and outflow velocity of the wind respectively, while the Thompson depth $\tau\sim1$ when $N_{\rm H}\sim10^{24}~{\rm cm^{-2}}$ as observed in the highly ionised wind.
\\
Thus the kinetic power (luminosity) of the wind is,

\begin{equation}
\dot{E}_{\rm w}=\frac{1}{2}\dot{M}_{\rm w} v_{\rm w}^{2}=\left ( \frac{v_{\rm w}}{2c} \right )L_{\rm flare},
\end{equation}
or integrating over time for the total energy:-

\begin{equation}
E_{\rm w} =\left ( \frac{v_{\rm w}}{2c} \right )E_{\rm flare}.
\end{equation}
It follows that an outflow velocity of $v_{\rm w}\sim0.25~c$, then $\dot{E}_{\rm w}\la0.15\dot{E}_{\rm flare}$, thus implying that we would expect only $\la15\%$ of the radiative power in the flare to be directly transferred to the wind.
\\
From the best fit to model to the eight slices, we estimated the luminosity of the flare between $1-1000$ Ryd to be $L_{\rm Flare(1-1000~Ryd)}\sim2\times10^{46}$ erg s$^{-1}$. 
\\
Now, the mass outflow rate of the wind is given by:-

\begin{equation}
\dot{M}_{\rm{w}} \sim \Omega m_pN_{\rm{H}}v_{\rm{w}}R_{\rm{in}},
\end{equation}
and based on the discussion in the previous sections we have adopted the following variables. From N15, we take the solid angle $\Omega=2\pi$ sr for the wide angle wind, the average column density measured between slices E - H $N_{\rm{H}} \sim 5\times10^{23}$ cm$^{-2}$, $v_{\rm{w}}\sim0.25~c$ and $R_{\rm{in}} \sim 100$ $R{\rm_g}\sim1.5\times10^{16}$ cm. Thus we estimated that $\dot{M}_{\rm{w}}\sim9\, M_\odot$ yr$^{-1}\sim0.4\dot{M}_{\rm{edd}}$ in good agreement with N15. Thus the kinetic luminosity of the outflow is $\sim1.5\times10^{46}$ erg s$^{-1}$ or $\sim0.1\, \textsc{L}_{\rm{edd}}$ for $M_{\rm BH}=10^{9}\odot{M}$. The duration of the wind in slices E - H is at least $\sim600$ ks, thus the mechanical energy deposited in the wind is at least $\sim10^{52}$ ergs. In order to be conservative we can assume that the flare has a duration corresponding to its onset to the end of the observational gap (beginning of slice E), i.e. $450$ ks in total. On this basis, by assuming that the flare persists at its maximum luminosity observed (i.e. $L_{\rm Flare(1-1000~Ryd)}\sim2\times10^{46}$ erg s$^{-1}$) it would impart a total radiative energy of $\sim8.5\times10^{51}$ erg. There are caveats attached to this scenario:- (i) by assuming that the extrapolated luminosity of the flare increased to such extend, $\sim 2\times10^{46}$ between $1-1000$ Ryd, we also rely on the assumption that the UV luminosity rises by the same amplitude to that of the X-rays which is rather unphysical; (ii) flare luminosities that remain constant for such an extended time-scale (i.e. $450$ ks) are usually undetected, thus such high and persisting value may be a considerable overestimation.
Even in the case, given only $\sim15\%$ of the radiative energy of $\sim8.5\times10^{51}$ erg s$^{-1}$ is deposited in the wind, it is unlikely that the subsequent outflow can be purely driven by radiation pressure by the flare alone. However magnetically driven outflows may provide an alternative mechanism for the initial driving for such powerful wind (Fukumura et al. 2015).

\subsection{Fractional Variability}

\begin{figure}[h!]
\includegraphics[width=60mm,height=50mm]{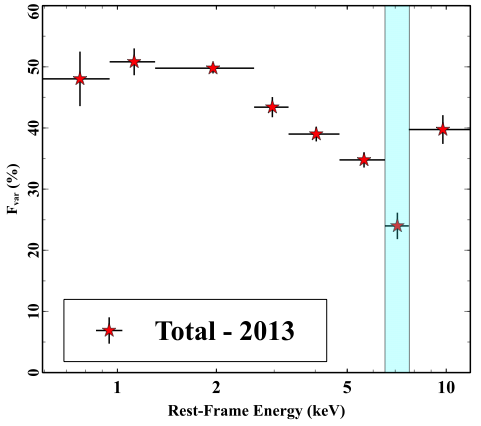}
\caption{Fractional variability from the 2013 observations. showing how the iron K emission band (rectangular area) centered at 7 keV is substantially less variable than the rest of the continuum. This may be associated with the reprocessed emission arising from the more distant material such as the wind.} 
\label{fig:fvar}
\end{figure}
We computed the fractional variability (F$_{\rm var}$) in different energy bands in the 2013 observation in PDS 456 using the (Vaughan et al. 2003) method. Looking at Fig.~\ref{fig:fvar}, what we see is a drastic decrease in F$_{\rm var}$ in the $6.5-7.7$ keV rest-frame energy band which is centred on the ionised Fe K emission. This may suggests that the Fe K emission line is less variable compared to the soft excess and the rest of the continuum or at the very least there is not substantial variability on the same short time-scale. 
\\
Furthermore it was worth investigating whether the iron K emission line responded to the variability of the continuum in a short time-scale. On this basis we tested two opposite scenarios:- in the first we kept the Fe K emission line normalisation fixed between the slices whereas in the second we allowed the emission line flux normalisation to vary between the slices \textit{in sync} with the continuum, so that the line EW is constant. As indicated from the F$_{\rm var}$ spectra, we can achieve a very good fit in the first constant flux scenario. However in the second scenario, a constant line EW produced a significantly worse fit compared to the constant line flux by $(\Delta\chi^{2}/{\nu}=100/8)$, thus indicating that there is no apparent short time-scale correlation between the Fe K line and the continuum flux.
\\ 
These results strongly suggest that on the short time-scales ($\sim100$ ks), the iron K emission is less variable than the continuum variations; this implies that the iron K emitting region may be larger than the typical continuum size inferred earlier of the order of $\sim6-20~R_{\rm g}$. Furthermore if the iron K emission arises from the disc, it may be located further out; on the other hand if the iron K emission is associated with the wind, as discussed in N15, it originates a  few $100~R_{g}$ further out from the black hole and therefore consistent with this argument.

\section{Conclusions}

In this work we have presented the results from the \textit{Suzaku} observing campaign ($\sim1$ Ms total duration) carried out in early 2013 of the nearby luminous quasar PDS 456. We investigated the broad-band continuum and absorption variability over the course of the low-flux 2013 observations, where the short-term spectral variability in PDS 456 was interpreted in terms of variable partially covering absorption 
\\
In the this scenario, we have found that the short-term spectral variability in the $\sim100$ ks time-scale may be due to variable inhomogeneous, neutral (or mildly ionised) clouds of gas that partially absorb the X-ray emission while crossing the LOS. Statistically speaking, the variable absorber model produces an excellent fit to the data. Furthermore we tested whether the spectral variability can be explained by variations in the intrinsic continuum only, where the partial covering absorbers are constant throughout the observation. We have found that this model seems to account for some parts of the observation; on the other hand this scenario fails to account for the spectral variability in particular before and after the flare. Generally speaking it follows that we cannot explain the overall short-term spectral variability behaviour without invoking a variable partial coverer combined, to a certain extent, with an intrinsically variable continuum. 
\\
On this basis, we have investigated in more detail the properties of the partial covering absorbers. We have found that these absorbers may be the least ionised component of the fast highly ionised outflowing wind with typical velocity of $\sim0.25$ c at the $99.9\%$ confidence level. We have shown that the short term variability of the iron K absorption may be attributed to the LOS variations in the column density (or ionisation). Through this variability, the size scale of the absorber is constrained to be $\sim20~R_{\rm g}$. Therefore following an obscuration event in the second half of the observation, i.e. in slice G, the size of the X-ray emitting region cannot be larger than $\sim20~R_{\rm g}$. In addition to this, we estimated the typical radial distance of the absorber from the black hole to be of the order of $\sim 200$ $R_{\rm g}$. 
\\
We have explored whether the radiation pressure imparted by the flare could produce enough kinetic power in the outflowing material to drive the wind. We have shown that the measured bolometric power arising from the flare was less than the mechanical power measured from the outflow by at least one order of magnitude, leading to the conclusion that another physical launching mechanism such as magnetically driven winds or the contribution of both. 
\\
Finally we calculated the fractional variability in the 2013 dataset as a function of energy. We found that the iron K emission band is substantially less variable than the rest of the continuum which might suggests that the iron K emission occurs from a much larger region than the continuum X-ray emission. This might be consistent with the emission arising from size scales above $\sim100~R_{\rm g}$.

\newpage%%%%%%%%%%%%%%%%%%%%%%%%%%%%%%%%%%%%%%%%%%%%%%%%%%%%%%


\begin{thebibliography}{}
	

\bibitem[Arnaud(1996)]{Arnaud96} Arnaud, K.~A.\ 1996, 
Astronomical Data Analysis Software and Systems V, 101, 17 



\bibitem[Ferrarese \& Merritt(2000)]{FM00} Ferrarese, L., \& Merritt, D.\ 2000, \apjl, 539, L9 


\bibitem[Fukumura et al.(2015)]{2015ApJ...805...17F} Fukumura, K., Tombesi, 
F., Kazanas, D., et al.\ 2015, \apj, 805, 17 



\bibitem[Gebhardt et al.(2000)]{Gebhardt00} Gebhardt, K., Bender, 
R., Bower, G., et al.\ 2000, \apjl, 539, L13 


	
	
\bibitem[Gofford et al.(2013)]{Gofford13} Gofford, J., Reeves, 
J.~N., Tombesi, F., et al.\ 2013, \mnras, 430, 60 
  


\bibitem[Hopkins 
\& Elvis(2010)]{HopkinsElvis10} Hopkins, P.~F., \& Elvis, M.\ 2010, \mnras, 401, 7 




\bibitem[King(2003)]{King03} King, A.\ 2003, \apjl, 596, L27


\bibitem[Matzeu et al.(2016)]{2016MNRAS.458.1311M} Matzeu, G.~A., Reeves, J.~N., Nardini, E., et al.\ 2016, \mnras, 458, 1311 



\bibitem[Nardini et al.(2015)]{2015Sci...347..860N} Nardini, E., Reeves, 
J.~N., Gofford, J., et al.\ 2015, Science, 347, 860 

   
 
\bibitem[Reeves 
\& Turner(2000)]{Reeves00} Reeves, J.~N., \& Turner, M.~J.~L.\ 2000, \mnras, 316, 234 
 

\bibitem[Reeves et al.(2003)]{Reeves03} Reeves, J.~N., O'Brien, 
P.~T., \& Ward, M.~J.\ 2003, \apjl, 593, L65 

  
 
\bibitem[Reeves et al.(2009)]{Reeves09} Reeves, J.~N., O'Brien, 
P.~T., Braito, V., et al.\ 2009, \apj, 701, 493 
 

\bibitem[Tombesi et al.(2010)]{Tombesi10} Tombesi, F., Sambruna, 
R.~M., Reeves, J.~N., et al.\ 2010, \apj, 719, 700 

 

%\bibitem[Tombesi et al.(2015)]{2015Natur.519..436T} Tombesi, F., Mel{\'e}ndez, M., Veilleux, S., et al.\ 2015, \nat, 519, 436 

 
 
\bibitem[Torres et al.(1997)]{Torres97} Torres, C.~A.~O., Quast, 
G.~R., Coziol, R., et al.\ 1997, \apjl, 488, L19 
  


\bibitem[Turner et al.(2011)]{2011ApJ...733...48T} Turner, T.~J., Miller, 
L., Kraemer, S.~B., \& Reeves, J.~N.\ 2011, \apj, 733, 48 



\bibitem[Vaughan et al.(2003)]{2003MNRAS.345.1271V} Vaughan, S., Edelson, 
R., Warwick, R.~S., \& Uttley, P.\ 2003, \mnras, 345, 1271 

  



\end{thebibliography}
\end{document}